\documentclass[superscriptaddress,aps,prd,nofootinbib,reprint]{revtex4-1}
\pdfoutput=1

\usepackage{graphicx}
\usepackage[colorlinks,allcolors=blue]{hyperref} 
\usepackage{amsmath,amssymb,bm}
\usepackage[T1]{fontenc}
\usepackage{mathtools}
\usepackage{multirow}
\usepackage{mdframed}

\begin{document}

\hfill TU-1243

\title{More is Different: Multi-Axion Dynamics Changes Topological Defect Evolution}

\author{Junseok Lee}
\affiliation{Department of Physics, Tohoku University, Sendai, Miyagi 980-8578, Japan} 
\author{Kai Murai}
\affiliation{Department of Physics, Tohoku University, Sendai, Miyagi 980-8578, Japan} 
\author{Fuminobu Takahashi}
\affiliation{Department of Physics, Tohoku University, Sendai, Miyagi 980-8578, Japan} 
\author{Wen Yin}
\affiliation{Department of Physics, Tokyo Metropolitan University, Tokyo 192-0397, Japan}

\begin{abstract}
We study topological defects in multi-axion models arising from multiple Peccei-Quinn (PQ) scalars. Using a simplified two-axion system, we reveal fundamental differences in the evolution of these defects compared to single-axion scenarios. This finding is particularly significant because, despite the fact that integrating out heavier axions reduces these models to an effective single PQ scalar theory at low energies, the actual physical behavior of topological defects differs markedly from single-axion predictions.
Unlike single-axion models where conventional cosmic strings form, multi-axion scenarios with post-inflationary or mixed initial conditions generically produce networks of strings interconnected by high-tension domain walls.
This results in a severe cosmological domain wall problem. We determine string-wall network instability conditions and discuss cosmological implications including the application to the QCD axion and gravitational wave generation. Our findings highlight that multi-axion dynamics can lead to qualitatively different outcomes for topological defects, challenging the conventional picture of cosmic evolution of topological defects based on single-axion models.
\end{abstract}

\maketitle

\section{Introduction}
The axion~\cite{Peccei:1977hh,Peccei:1977ur,Weinberg:1977ma,Wilczek:1977pj}, a Nambu-Goldstone (NG) boson arising from the spontaneous symmetry breaking (SSB) of the global U(1) Peccei-Quinn (PQ) symmetry, is well-known as a promising dark matter candidate due to its naturally light mass and weak interactions. See  Refs.~\cite{Kim:2008hd,Arias:2012az,Marsh:2015xka,DiLuzio:2020wdo,OHare:2024nmr} for reviews. Axions can be produced through the misalignment mechanism~\cite{Preskill:1982cy,Abbott:1982af,Dine:1982ah} or the decay of topological defects~\cite{Sikivie:1982qv,Vilenkin:1982ks,Harari:1987ht,Davis:1986xc}.

In recent years, there has been a significant advancement in the precision of lattice-based numerical simulations focusing on the formation of axion strings and their subsequent cosmic evolution~\cite{Saikawa:2024bta,Kim:2024wku,Buschmann:2024bfj}. These simulations typically introduce a single PQ scalar field to spontaneously break the global U(1) PQ symmetry. This approach is justified by the assumption that once axion strings form, the details of the PQ sector become UV physics confined to the string core, thus not affecting the global string evolution. Consequently, numerical calculations can be performed with only one PQ scalar and thus one axion.

In the case of a single PQ scalar, topological defects arising from post-inflationary initial conditions generally produce more axions than in the pre-inflationary scenario, given the same decay constant. Additionally, domain walls (DWs) cause the cosmological DW problem~\cite{Zeldovich:1974uw} unless the DW number equals unity. Therefore,  changing from post-inflationary to pre-inflationary initial conditions generally mitigates the cosmological impact of axions. 

While these single-axion models have provided valuable insights, they do not capture the full complexity of potential PQ sector structures.  
In fact, it was already pointed out in Ref.\cite{Higaki:2016jjh} that when the PQ sector has a highly complex structure, particularly in cases where the number of axions $N_{\rm axion}$ is large, it becomes extremely difficult to form conventional axion strings, and the UV physics of the PQ sector can play a crucial cosmological role.
This study applies the clockwork (or alignment) mechanism~\cite{Kim:2004rp,Higaki:2014pja,Choi:2014rja,Choi:2015fiu,Kaplan:2015fuy,Giudice:2016yja}  to the QCD axion following Ref.~\cite{Higaki:2015jag}, and discusses the types of topological defects that arise in the clockwork mechanism (see also Ref.~\cite{Long:2018nsl}).

The results reveal that the structure within the core of axion strings at low energies consists of an intricate assembly of strings and DWs, termed a ``string bundle''~\cite{Higaki:2016jjh}. This string bundle contains $N_{\rm axion}$ types of strings, with an exponential hierarchy required among their numbers. Consequently, forming such a string bundle from individual strings that evolve following the scaling solution would take an exponentially long time. As a result, in realistic cosmic evolution, it becomes impossible to form a string bundle, and the UV physics that should be confined to the axion string core spreads throughout the universe. Specifically, DWs with large tension are considered to extend across the cosmic scales. As stressed in Ref.~\cite{Higaki:2016jjh}, the decay of these high-tension DWs can generate extremely strong gravitational waves, which can be probed by observations such as pulsar timing arrays. Interestingly, such axion DWs in multi-axion systems potentially explain the recent hint for stochastic gravitational waves found by the pulsar timing arrays such as NANOGrav~\cite{NANOGrav:2023gor,Antoniadis:2023ott,Reardon:2023gzh,Xu:2023wog}. See Refs.~\cite{Kitajima:2023cek,Ferreira:2024eru,Dankovsky:2024zvs,Blasi:2023sej} for the domain-wall explanation of the NANOGrav signal. 

In this paper, we generalize the multi-axion scenario beyond the clockwork QCD axion model considered in Ref.~\cite{Higaki:2016jjh}. We demonstrate that even in a simplified two-axion model, stable high-tension DW networks can form, similar to those found in the clockwork model~\cite{Higaki:2016jjh}. This occurs with post-inflationary initial conditions, without requiring the numerous axions characteristic of the clockwork mechanism. Furthermore, we show that string bundles similarly fail to form with mixed pre- and post-inflationary initial conditions. These stable string-wall networks lead to the cosmological DW problem. To keep our analysis tractable, we focus on a two-axion setup with mixing, revealing that the DW problem can arise in much simpler multi-axion scenarios than previously thought. We systematically explore the conditions under which this string-wall network becomes unstable. Furthermore, we discuss cosmological implications of this multi-PQ scalar sector, including the potential generation of significant gravitational wave signals from the decay of these high-tension DWs.

Finally, we comment on the relevant literature on multiple axions, and mixed pre- and post-inflationary initial conditions. 
The topological defects in multiple axion models were studied in Refs.~\cite{Higaki:2016jjh,Long:2018nsl,Hiramatsu:2019tua,Eto:2023aqr,Lee:2024xjb}. 
In our recent work~\cite{Lee:2024xjb}, we studied a two-axion system with mixed post- and pre-inflationary initial conditions. We showed that the string-wall network associated with one axion with the post-inflationary initial condition can induce a domain-wall-like structure for the other axion with the pre-inflationary initial condition. As we will show later, these ``induced domain walls'' play a crucial role in multi-axion systems, particularly in understanding the stability of the string-wall network.
In this paper, when we consider pre-inflationary conditions, we focus on scenarios where the corresponding axion strings have been sufficiently diluted by inflation to be effectively absent from the observable universe.  It is worth noting, however,  that there are scenarios considering a very sparse distribution of strings. Such scenarios can arise if SSB occurs during or near the end of inflation~\cite{Lazarides:1984pq,Shafi:1984tt,Vishniac:1986sk,Yokoyama:1988zza}, or if late-time inflation, such as thermal inflation~\cite{Yamamoto:1985rd,Lazarides:1985ja,Lyth:1995hj,Lyth:1995ka}, is considered. Recently, Ref.~\cite{Bao:2024bws} examined a system of two mixing axions, applying post-inflationary conditions to one and this highly diluted string distribution to the other, focusing on the generation of gravitational waves. In the context of the string axiverse, Ref.~\cite{Benabou:2023npn} studied axion strings and DWs in a two-axion system, similar to the pre-post scenario we will discuss later. Their work, however, did not address the induced DWs. We will show that the induced DWs play a crucial role in determining the network's stability.

The structure of this paper is as follows: we first summarize our main results at the end of this section. 
In Sec.~\ref{sec: model} we give the set-up for the two-axion system with mixing.
In Sec.~\ref{sec: DW} we discuss the formation and time evolution of DWs in the two-axion models.
We consider the DW formation due to one potential term in Sec.~\ref{subsec: first DW} and that due to another potential term in Sec.~\ref{subsec: second DW}.
We briefly mention special cases in Sec.~\ref{subsec: n1orn2=0} and summarize the classification of scenarios based on the DW numbers in Sec.~\ref{subsec: summary classification}.
In Sec.~\ref{sec: cosmology} we discuss cosmological implications of topological defects in multi-axion scenarios. 
The last section is devoted to discussion and conclusions.

\subsection{Overview of results}
Let us summarize our key findings. In the following sections, we will examine a system of multiple axions arising from multiple PQ scalars, where the lightest axion is significantly lighter than the others. By integrating out all the heavier axions, this system reduces to a single-axion system at low energies, for which a conventional axionic cosmic string is defined. Here, all the intricate high-energy physics of the heavier axions is confined to the string core. As a result, in the system of multiple PQ scalars,  the conventional axionic string is actually a bundle of strings, termed a ``string bundle''~\cite{Higaki:2016jjh}. Our key findings regarding the system of multiple PQ scalars are as follows:
\begin{itemize}
\item In the post-inflationary scenario, the conventional axionic strings in the low-energy effective theory do not necessarily form, even in a two-axion system. Instead, a string-wall system of heavy axions could persist, leading to the cosmological DW problem.
\item A stable string-wall system of heavy axions also appears when mixed pre- and post-inflationary initial conditions are imposed on the multiple PQ scalars.
\item For successful cosmology, this string-wall system must collapse, resulting in the emission of a significant amount of gravitational waves during its decay.
\end{itemize}
Thus, the multiple-axion scenario presents a dramatically different evolution of topological defects compared to conventional single-axion models. In particular, the aforementioned string-wall system is analogous to that found in the clockwork axion model~\cite{Higaki:2016jjh}. However, we demonstrate that this phenomenon occurs within a much broader framework of multiple axions.

\section{Two-axion system with mixing}
\label{sec: model}

We consider a simple setup with two axions, $\phi_1$ and $\phi_2$. These axions appear in the phases of complex PQ scalar fields $\Phi_1$ and $\Phi_2$, respectively:
\begin{align}
\label{eq:PQscalar}
    \Phi_j = \frac{f_j}{\sqrt{2}}e^{i \frac{\phi_j}{f_j}},
\end{align}
where $j = 1,2$, and $f_j$ denotes the decay constant of $\phi_j$. Thus,
the axions, $\phi_1$ and $\phi_2$, are the NG bosons of $U(1)_{\phi_1} \times U(1)_{\phi_2}$.
For later use, we define the angles, $\theta_i \equiv \phi_i/f_i$.
Except when considering axion strings, we neglect the radial degrees of freedom, assuming that they are much heavier than the axions. It is worth noting that when imposing pre-inflationary initial conditions, such UV completion is not necessary; we could instead consider axions originating from gauge fields in extra dimensions, as in string axion models, where there is no symmetry restoration. From Eq.~\eqref{eq:PQscalar}, it is evident that $\phi_1$ and $\phi_2$ have periodicities of $2\pi f_1$ and $2 \pi f_2$, respectively.

Let us suppose that $\Phi_j$ is initially at the origin. As $\Phi_j$ develops a non-zero vacuum expectation value (VEV), it spontaneously breaks the corresponding U(1) symmetry, resulting in the formation of cosmic strings associated with $\phi_j$. In the case of pre-inflationary initial conditions, these cosmic strings are effectively diluted due to the subsequent inflationary expansion. Later, when the axion begins to oscillate, DWs emerge.\footnote{
In this paper, we do not consider the DW formation due to quantum fluctuations in the pre-inflationary scenario~\cite{Linde:1990yj,Lyth:1991ub,Takahashi:2020tqv,Gonzalez:2022mcx}, where there are no strings attached to the walls.
}
We will see that how the DWs are connected to strings, as well as the subsequent dynamics of the string-wall network crucially depends on the mixing between the axions and the initial conditions.

We consider the effective Lagrangian for two axions, $\phi_1$ and $\phi_2$, given by
\begin{align}
\mathcal{L}
=
\frac{1}{2} \partial_\mu \phi_1 \partial^\mu \phi_1
+ \frac{1}{2} \partial_\mu \phi_2 \partial^\mu \phi_2
- V(\phi_1, \phi_2)
\ ,
\end{align}
where the potential $V(\phi_1, \phi_2)$ consists of two terms:
\begin{align}
V(\phi_1, \phi_2) = V_1(\phi_1, \phi_2) + V_2(\phi_1, \phi_2)
\ .
\label{eq: potential}
\end{align}
These potential terms are defined as follows:
\begin{align}
V_1(\phi_1, \phi_2) &=
\Lambda^4 \left[
1 - \cos \left( n_1 \frac{\phi_1}{f_1} + n_2 \frac{\phi_2}{f_2} \right)
\right]
\ , \\
V_2(\phi_1, \phi_2) &=
\Lambda^{\prime 4} \left[
1 - \cos \left( n'_1\frac{\phi_1}{f_1} + n'_2\frac{\phi_2}{f_2}
+ \alpha \right)
\right]
\ .
\end{align}
Here, $\Lambda$ and $\Lambda'$ are the energy scales of the potential terms,
$n_1, n_2, n_1'$, and $n_2'$ are integers known as DW numbers, and $\alpha$ represents a phase constant. Let us define $\theta_1 \equiv \phi_1/f_1$ and $\theta_2 \equiv \phi_2/f_2$ for later use.
In our analysis, we consider cases where either all four integers $n_1$, $n_2$, $n_1'$, and $n_2'$ are non-zero, or at most one of them is zero. We exclude scenarios where two or more of these integers are zero, as such cases would reduce to either a situation without mixing or one involving only a single potential. 
If $D \equiv |n_1 n_2' - n_2 n_1'| \neq 0$, we can set $\alpha = 0$ by shifting $\phi_1$ and/or $\phi_2$. However, for $D = 0$, the phase constant $\alpha$ is generically non-zero.

For simplicity, we consider a scenario where the axions begin oscillating first due to $V_1$ and then, much later in the cosmic evolution, due to $V_2$. This implies a hierarchy between the two mass eigenvalues, $$m_\mathrm{heavy} \gg m_\mathrm{light},$$ where $m_\mathrm{heavy}$ and $m_\mathrm{light}$ represent the heavy and light mass eigenvalues at the minimum, respectively. 
The DWs appear when the axions begin oscillating, if the post-inflationary initial condition is assumed. To simplify our analysis, we also assume a hierarchy in the DW tensions: the tension of DWs due to $V_1$ is much larger than that of DWs due to $V_2$.

Given these conditions, we define the heavy and light axions, $\phi_\mathrm{heavy}$ and $\phi_\mathrm{light}$, as:
\begin{align}
\phi_\mathrm{heavy} &= F \left(n_1 \frac{\phi_1}{f_1} + n_2 \frac{\phi_2}{f_2} \right),\\
\phi_\mathrm{light} & = F \left(-n_2 \frac{\phi_1}{f_2} + n_1 \frac{\phi_2}{f_1} \right),
\end{align}
where
\begin{align}
F &= \frac{f_1 f_2}{\sqrt{(n_1 f_2)^2 + (n_2 f_1)^2}}.
\end{align}
It is important to note that $\phi_\mathrm{heavy}$ and $\phi_\mathrm{light}$ are canonically normalized mass eigenstates at the minimum of $V_1$ in the absence of $V_2$, and remain approximate mass eigenstates as long as $V_2$ is sufficiently small.

In the absence of the axion potential $V(\phi_1, \phi_2)$, the system originally possesses a $U(1)_{\phi_1} \times U(1)_{\phi_2}$ symmetry, under which $\theta_1$ and $\theta_2$ transform, respectively, with a period of $2\pi$. Since a linear combination of $U(1)_{\phi_1}$ and $U(1)_{\phi_2}$ is broken by $V_1$, it is convenient to define a new basis $U(1)_H \times U(1)_L$, under which $\theta_H$ and $\theta_L$ transform, respectively, with a period of $2\pi$. Here, $\theta_H$ and $\theta_L$ are defined as:
\begin{align}
\theta_H & \equiv \frac{d}{n_1^2 + n_2^2}(n_1 \theta_1 + n_2 \theta_2),\\
\theta_L &\equiv \frac{d}{n_1^2 + n_2^2} (-n_2 \theta_1 + n_1\theta_2),
\end{align}
with $d$ being the greatest common divisor of $(|n_1|,|n_2|)$.%
\footnote{Here, we define $d = n_2$ for $n_1 = 0$ and $d = n_1 $ for $n_2 =0$.}
As we will see,  $\theta_L$ changes by $2\pi$ while $\theta_H$ stays constant as one goes around the string bundle.
Note that $\theta_L$ is not proportional to $\phi_\mathrm{light}$ unless $f_1 = f_2$ while we define $\theta_H$ to be proportional to $\phi_\mathrm{heavy}$. This is because we obtain $(\phi_\mathrm{heavy}, \phi_\mathrm{light})$ and $(\theta_H, \theta_L)$ from orthogonal transformations of $(\phi_1, \phi_2)$ and $(\theta_1, \theta_2)$, respectively.

When we apply this system to the QCD axion, the light mass eigenstate $\phi_\mathrm{light}$ is to become the QCD axion, and $V_2$ arises from non-perturbative effects of QCD. 
In this case, $U(1)_{L}$ is identified with the usual $U(1)_{\rm PQ}$ symmetry.
Thus, the two-axion model can be thought of as a toy UV completion of the QCD axion.
To make our analysis tractable, however, we simply assume both $\Lambda$ and $\Lambda'$ are time-independent, and our analysis can be straightforwardly applied to the temperature-dependent axion potentials. We comment on the application to the QCD axion later in this paper.

In multiple axion models, the axion dynamics can be classified
based on the timing of the spontaneous symmetry breaking as in the single axion models. With two axions, we can categorize the situations into three scenarios:
\begin{enumerate}
\item ``Post-post scenario'': Both axions have post-inflationary initial conditions
\item ``Pre-post scenario'': One axion has a pre-inflationary initial condition, while the other has a post-inflationary initial condition
\item ``Pre-pre scenario'': Both axions have pre-inflationary initial conditions  
\end{enumerate}
We will use these terms throughout the following discussion.

\section{Formation and evolution of domain walls}
\label{sec: DW}

In the pre-pre scenario, both axions $\phi_1$ and $\phi_2$ have almost uniform field values, and no cosmic strings are formed. Thus, DWs are not formed either, and the evolution of the axions is described by the background dynamics. On the other hand, in the post-post and pre-post scenarios, cosmic strings are formed associated with the axion that emerges at the SSB after inflation. Subsequently, DWs are formed corresponding to the potential that makes the axion oscillate. In the following, we discuss the formation and evolution of DWs due to $V_1$ and $V_2$ in order. For the moment, we assume that both $n_1$ and $n_2$ are nonzero and briefly comment on the case where either $n_1$ or $n_2$ vanishes later.

\subsection{First DW formation due to \texorpdfstring{$V_1$}{}}
\label{subsec: first DW}

First, we discuss the DW formation due to $V_1$, neglecting $V_2$. 
From the point of view of symmetry groups, the system initially has a $U(1)_H \times U(1)_L$ symmetry, and $U(1)_H$ is explicitly broken to $\mathbb{Z}_{(n_1^2+n_2^2)/d}$ in the presence of $V_1$, while $U(1)_L$ remains intact.
Although $\phi_\mathrm{heavy}$ becomes massive due to $V_1$, it cannot simply be integrated out even at low energies because cosmic strings associated with $\phi_1$ and/or $\phi_2$ are formed in the post-post and pre-post scenarios.
To integrate out $\phi_\mathrm{heavy}$, we need to confine it to the core of string bundles. However, as we will see below, this is typically not possible in multiple axion models.

\subsubsection{Post-post scenario}
\label{subsubsec: first post-post}

We begin with the post-post scenario. First, cosmic strings of $\phi_1$ and $\phi_2$ are formed due to the SSB of $U(1)_{\phi_1}$ and $U(1)_{\phi_2}$, respectively. Shortly thereafter, strings of these two types respectively approach a scaling solution where ${\cal O}(1)$ strings are present within each Hubble horizon. 
As the axions begin to oscillate due to $V_1$, DWs appear. Specifically, $|n_1|$ and $|n_2|$ 
DWs are attached to the string of $\phi_1$ and $\phi_2$, respectively.
Then, the string-wall network is formed, where cosmic strings of $\phi_1$ and $\phi_2$ are connected by DWs of $V_1$.

\begin{figure}[!t]
    \begin{center}
    \centering
    \begin{minipage}[b]{0.25\textwidth}
        \centering
        \includegraphics[width=\textwidth]{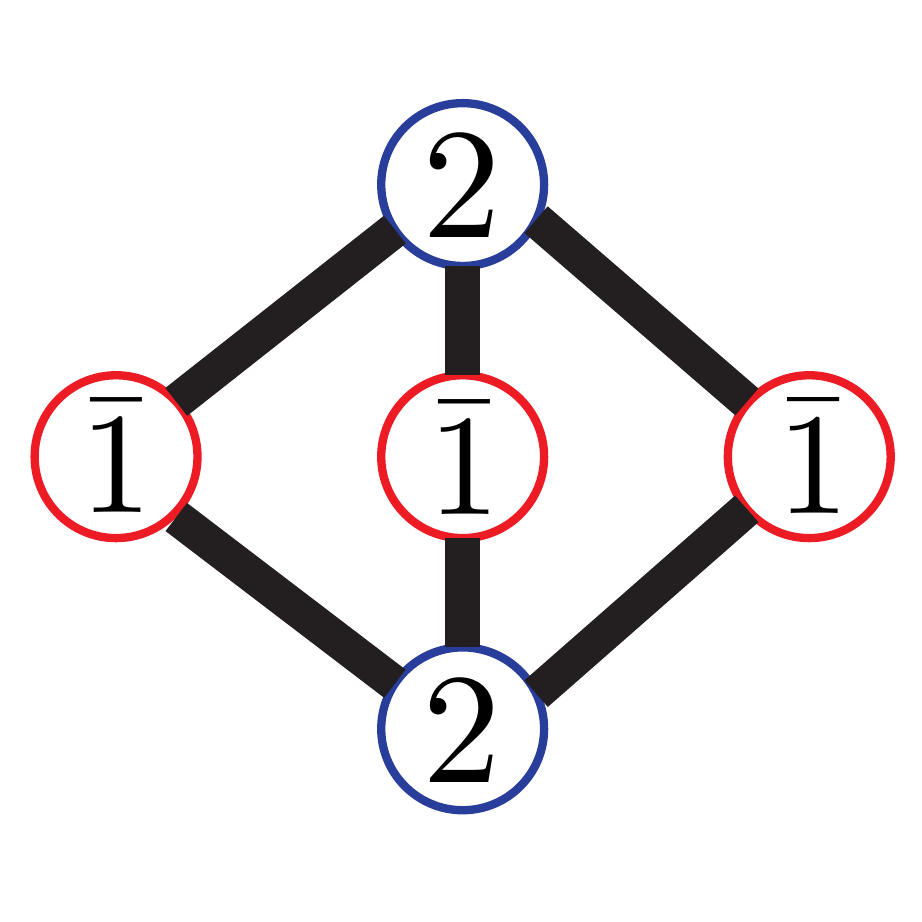}
        \makebox[1.\textwidth]{(a)}
    \end{minipage}
    \hfill
    \begin{minipage}[b]{0.45\textwidth}
        \centering
        \includegraphics[width=\textwidth]{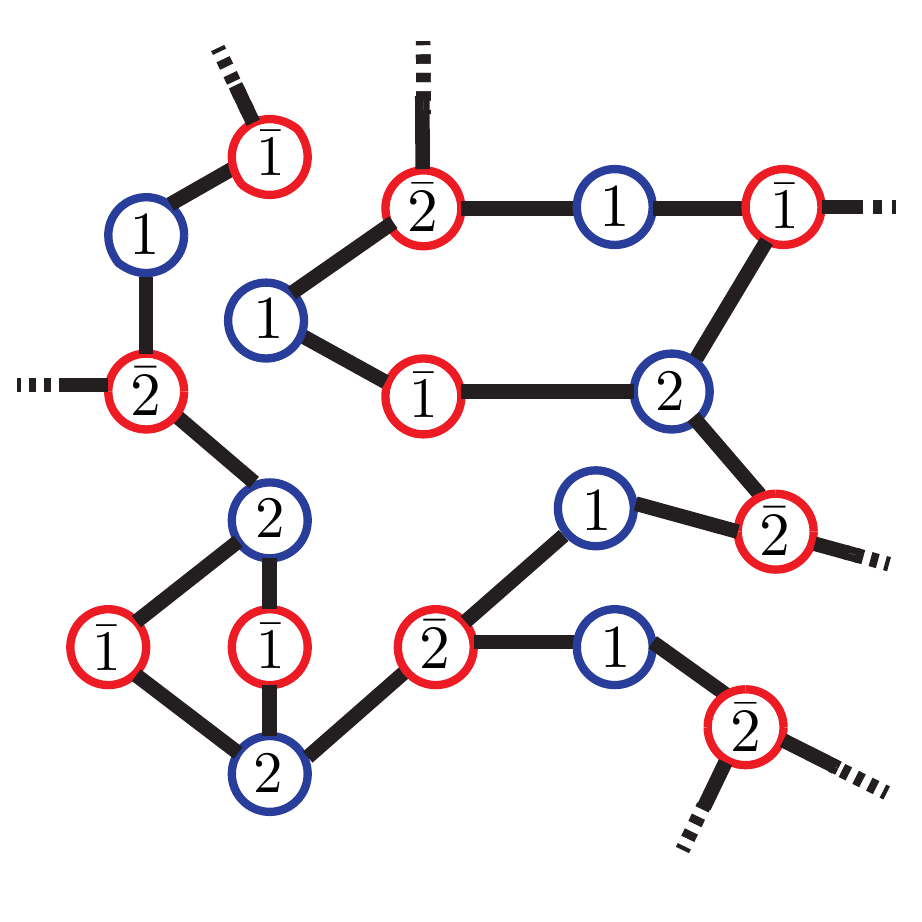}
        \makebox[1.\textwidth]{(b)}
    \end{minipage}
    \end{center}
    \caption{%
Schematic representations of (a) a string bundle and (b) a DW network, with $n_1 = 2$ and $n_2 = 3$.
The circles represent the strings (blue) and anti-strings (red) of $\phi_1$ and $\phi_2$. The thick lines connecting them represent DWs due to $V_1$.
    }
    \label{fig: string bundle} 
\end{figure}

Since we neglect $V_2$ for the moment, the lighter axion, $\phi_\mathrm{light}$, is massless, and we can construct a string bundle of $\theta_L$ to which no DWs are attached. Such a string bundle is composed of $n_2/d$ anti-strings of $\phi_1$ and $n_1/d$ strings of $\phi_2$. See the upper panel of Fig.~\ref{fig: string bundle} for the schematic representation of a string bundle with $n_1 = 2$ and $n_2 = 3$,
where the anti-strings of $\phi_1$ are represented by $\bar{1}$, and the strings of $\phi_2$ are represented by $2$. Here, (anti-)strings are defined such that the phase $\theta_1$ or $\theta_2$ increases (decreases) in a counterclockwise direction when we consider a plane intersecting the strings. As one can see from the upper panel of Fig.~\ref{fig: string bundle}, $\theta_L$ changes by $2\pi$ around the string bundle, since $\theta_1$ and $\theta_2$ change by $-2\pi n_2/d$ and $2\pi n_1/d$, respectively. 
If all the strings form such string bundles or similar isolated objects, the DWs due to $V_1$ are confined to the cores of these string bundles, and only these bundles remain in the universe. Note that, at low energies, these string bundles behave similarly to ordinary axion strings associated with the lighter axion, $\phi_\mathrm{light}$. On the other hand, a substantial fraction of strings could remain outside the string bundles, if the strings do not perfectly align to form such decoupled bundles. In this case, DWs stretch between these unbundled strings, forming an interconnected network. See the lower panel of Fig.~\ref{fig: string bundle} for the schematic representation of a remaining string-wall network. As a result, the DW network persists throughout the universe, potentially leading to the DW problem.

Whether all strings form the bundles will depend on the DW numbers, $n_1$ and $n_2$.
Here we assume that $f_1$ and $f_2$ are not hierarchical so that the decay constant of the heavy axion, $F$, is of the same order as $f_{1,2}$ unless large $n_{1,2}$ is assumed.
If either $|n_1|$ or $|n_2|$ equals unity, the corresponding strings become endpoints of the DW network, and the strings likely form the bundles.
Thus, we expect the DW network to disappear when either $|n_1| = 1$ and $|n_2| = {\cal O}(1)$, or $|n_1| = {\cal O}(1)$ and $|n_2| = 1$.\footnote{%
In Ref.~\cite{Higaki:2016jjh}, the case of three axions with two axion potentials was considered, where one of the axions has $n_1 = 1$. In this case, it was shown that a stable DW network remains.
Note also that it remains uncertain whether string bundles will form when one of the DW numbers is significantly larger than ${\cal O}(1)$, even if the other equals $\pm 1$. 
In Ref.~\cite{Eto:2023aqr}, the case of $n_1 = 2$ and $n_2 = -1$ is studied assuming a hierarchical symmetry breaking associated with two axions.
}
On the other hand, if $|n_1|\geq 2$ and $|n_2| \geq 2$, we expect that the string-wall network survives.
This expectation stems from the observation for $n_1 = n_2 = 2$.
In this case, the structure of the string-wall network is similar to that in the single-axion model with a DW number of $2$, except for the existence of two species of cosmic strings.
From numerical simulations, it is known that the DW network is stable in the single-axion case, which implies the stability of the DW network in the two-axion case.
For larger $|n_1|$ and/or $|n_2|$, the structure of the string-wall network becomes more complicated, which makes our expectation more convincing.

\begin{figure}[!t]
    \begin{center}  
        \includegraphics[width=0.45\textwidth]{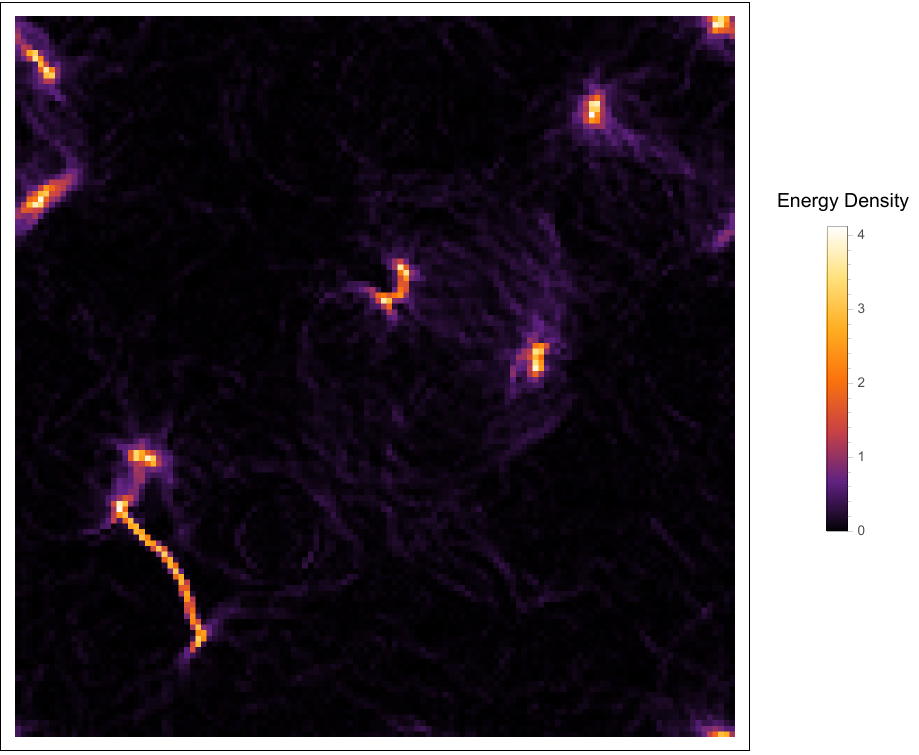}
        \\
    \includegraphics[width=0.45\textwidth]{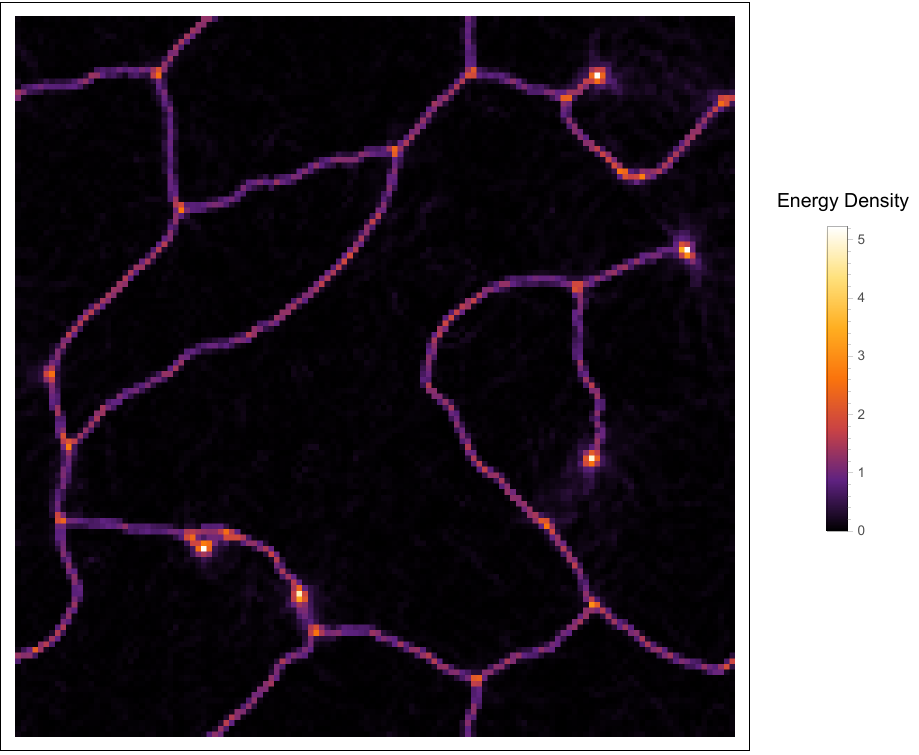}
    \end{center}
    \caption{
 The snapshots of the energy density at the time $t = 300/\sqrt{\lambda}\eta$ for $(n_1,n_2) = (1,2)$ (upper panel) and $(3,2)$ (lower panel) are shown. The energy density is normalized by $\lambda \eta^4$. In the upper panel, several string bundles, each composed of three strings, are formed. In contrast, a string-wall network is formed in the lower panel. In both panels, one side of the box is about the size of two Hubble horizons.
    }
    \label{fig: 2Dnumerical simulation} 
\end{figure}

To validate our expectations, we numerically investigate the evolution of the string-wall network through lattice simulations.
To simulate the dynamics of the strings, we consider $\Phi_i$ with the Lagrangian of 
\begin{align}
    \mathcal{L}_\Phi 
    =&
    \sum_{i=1}^2 \left[ 
        |\partial \Phi_i|^2 
        - \frac{\lambda_i}{4} \left( |\Phi_i|^2 - \frac{f_i^2}{2} \right)^2
    \right]
    \nonumber \\
    &+ \left( 
        \epsilon \Phi_1^{n_1} \Phi_2^{n_2}
        + \mathrm{h.c.}
    \right)
    \ ,
\end{align}
where $\lambda_1$, $\lambda_2$, and $\epsilon$ are real constants.
For simplicity, we assume $f_1 = f_2 \equiv \sqrt{2}\eta$ and $\lambda_1 = \lambda_2 \equiv \lambda$, and set $\epsilon = 0.1 \lambda \eta^{4-n_1-n_2}/(n_1^2+ n_2^2)$, which corresponds to $m_\mathrm{heavy}^2 = 0.1 \lambda \eta^2$ with $m_\mathrm{heavy}$ being the mass of $\phi_\mathrm{heavy}$.
We have performed $(2+1)$-dimensional lattice simulations with $2048^2$ grids for $(n_1,n_2) = (1,2)$ and $(3,2)$, assuming the radiation-dominated universe. See Fig.~\ref{fig: 2Dnumerical simulation}. In the upper panel with $(n_1,n_2) = (1,2)$, several string bundles, each composed of two $\phi_1$ strings and one $\phi_2$ string, are formed. In contrast, in the lower panel with $(n_1,n_2)=(3,2)$, the string-wall network is formed as expected.

We have considered a two-axion system without a hierarchy in the decay constants to discuss the formation of string bundles.
Note that if a hierarchy between the decay constants exists, the DWs take longer to shrink, as we show here.
For the case where $|n_1| = 1$ and $|n_2| = \mathcal{O}(1)$ but not equal to 1 (or vice versa), the DWs bounded by strings contract soon {after} the DW tension force becomes comparable to the string tension force, at which point the Hubble parameter is given by
\begin{align}
    H \sim \frac{m_{\rm heavy} F^2}{f_i^2}\,,
\end{align}
where we choose $i\,(=1,2)$ such that the strings associated with $\phi_i$ bound the DWs.
Assuming $f_1 \ll f_2$ for instance, $F \simeq f_1/n_1$ is much smaller than $f_2$.
If $|n_1| = 1$ and $|n_2| > 1$, the strings associated with $\phi_2$ dominate the dynamics of the string-wall network, resulting in a delayed bundle formation by a factor of $(f_2/f_1)^2$.
Since the strings continue to dominate the string-wall network, the delayed formation does not lead to the DW problem. See also Ref.~\cite{Eto:2023aqr} for the case of hierarchical decay constants.

\subsubsection{Pre-post scenario}
\label{subsubsec: first pre-post}

In the pre-post scenario, we consider, without loss of generality, the pre-inflationary scenario for $\phi_1$ and the post-inflationary scenario for $\phi_2$. Then, cosmic strings of $\phi_2$ are formed while $\phi_1$ is almost spatially homogeneous. This is equivalent to diluting away only the cosmic strings of $\phi_1$. Thus, it becomes impossible to form string bundles that would contain $|n_2|/d$ strings of $\phi_1$. So, we are left only with cosmic strings of $\phi_2$.

In this case, the cosmic strings of $\phi_2$ evolve in the same way as in the single-axion case with the DW number of $n_2$.
If $|n_2| = 1$, the string-wall network rapidly decays after the formation of DWs, and no topological defects survive.
If $|n_2| \geq 2$, on the other hand, the network of cosmic strings of $\phi_2$ and the DWs of $V_1$ attached to them is stable.

\begin{figure}[!t]
    \begin{center}  
        \includegraphics[width=0.45\textwidth]{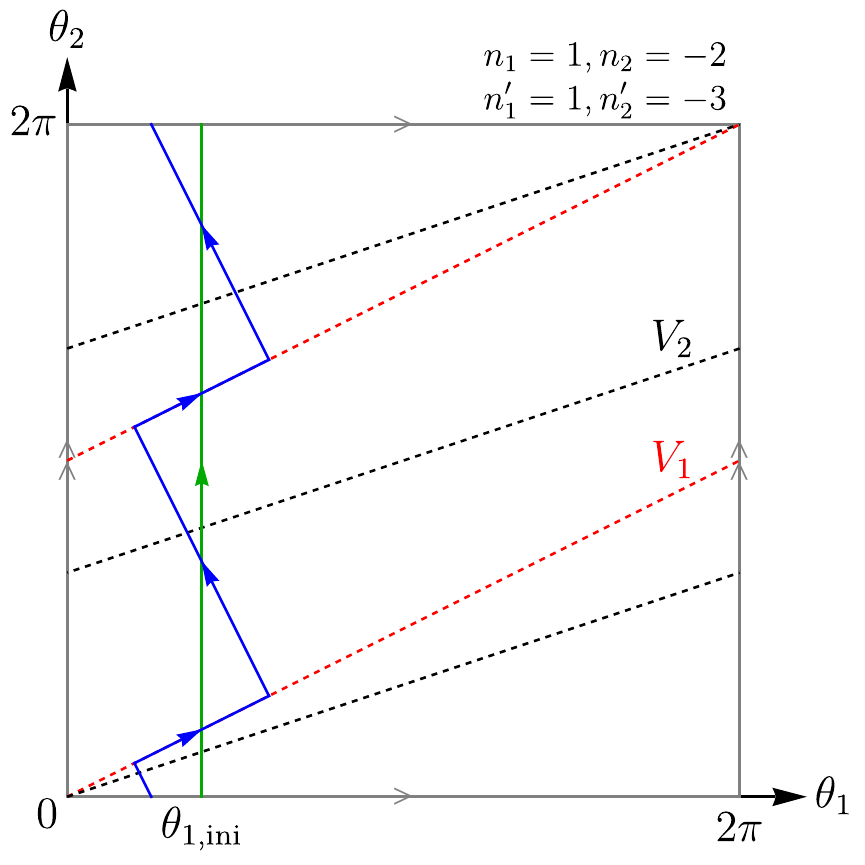}
        \\
        \includegraphics[width=0.45\textwidth]{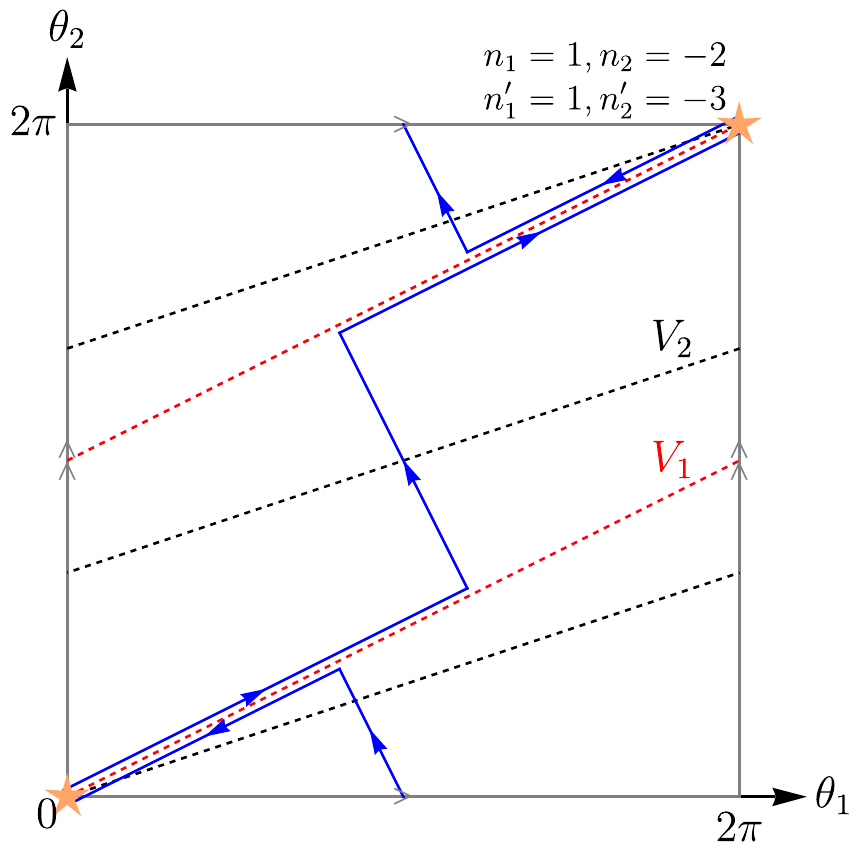}
    \end{center}
    \caption{%
        \textbf{Upper panel:} Field configuration of the axions around the cosmic string of $\phi_2$ in the pre-post scenario with $(n_1, n_2) = (1,-2)$, neglecting the effect of $V_2$.
        The winding numbers of $\phi_1$ and $\phi_2$ are $0$ and $1$, respectively, reflecting their initial conditions. 
        The green and blue lines represent the initial configuration and the DW configuration due to $V_1$, respectively.
        The red and black dashed lines are the minima of $V_1$ and $V_2$, respectively.
        Note that this figure does not depict exact angles between each segment of the configuration. The actual angles depend on the ratio of $f_1$ to $f_2$.
        \textbf{Lower panel:} Same as the upper panel except for the inclusion of $V_2$ with $(n_1', n_2') = (1, -3)$. The blue line segments perpendicular to and along the red-dashed line represent the configuration of the DWs due to $V_1$ and the induced DWs due to $V_2$, respectively. The orange star represents the global minimum of the total potential, which is the bulk domains between the DWs.
    }
    \label{fig: pre-post} 
\end{figure}

Here, let us fix $(n_1, n_2) = (1, -2)$ for concreteness.
Each string of $\phi_2$ has two DWs, and the axions settle down to one of the minima of $V_1$ in each domain.
Then, the DW network due to $V_1$ is stable as discussed above.
We show the trajectory of the axions around a cosmic string in the upper panel of Fig.~\ref{fig: pre-post}. The minimum of $V_1$ is shown by the red-dashed line.
Until the axions begin oscillating due to $V_1$, $\theta_1$ maintains its initial value, $\theta_{1,\mathrm{ini}}$, and the field trajectory around the string of $\phi_2$ is described by the vertical green line. Then, when the axion begins oscillating due to $V_1$, the trajectory is deformed to minimize the energy of the system, as shown by the blue line.
Since the DW tension becomes smaller for a shorter field distance between the different minima of $V_1$, the trajectory of the DW is perpendicular to the minimum of $V_1$ in the $(\phi_1,\phi_2)$ plane.
The blue line along the minimum of $V_1$  corresponds to the bulk of each domain.

Thus, by considering the mixed pre-post inflationary scenario, the strings of $\phi_2$ form a stable string-wall network if $|n_2| \geq 2$, even if $n_1 = 1$.
Note that for $n_1 = 1$ and $n_2 = -2$, string bundles are likely to form in the post-post scenario. Thus, unlike the single-axion scenarios, changing the initial condition from the post-inflationary condition to the pre-inflationary condition does not necessarily make the cosmological impact of the axion topological defects mild. 

\subsection{Second DW formation due to \texorpdfstring{$V_2$}{}}
\label{subsec: second DW}

Next, we discuss the DWs due to $V_2$.
If the DW network due to $V_1$ persists after $V_2$ becomes relevant, $V_2$ can induce additional DWs along the DWs of $V_1$, which we call induced DWs~\cite{Lee:2024xjb}.
The emergence of induced DWs is more evident in the pre-post scenario, so we discuss this scenario first, followed by the post-post scenario.

\subsubsection{Pre-post scenario}
\label{subsubsec: Second pre-post}

We consider the case of $|n_2| \geq 2$ so that the DW number of $\phi_2$ strings equals $|n_2| \geq 2$ after $V_1$ becomes relevant.
We continue to work on the case of $(n_1, n_2) = (1, -2)$ as a representative example in this case.
The subsequent evolution of the string-wall network depends on $(n_1', n_2')$.
We classify the cases into $D \neq 0$ and $D = 0$. We remind that $D$ was defined as $D \equiv |n_1 n_2' - n_2 n_1'|$.

\begin{figure}[!t]
    \begin{center}  
        \includegraphics[width=0.45\textwidth]{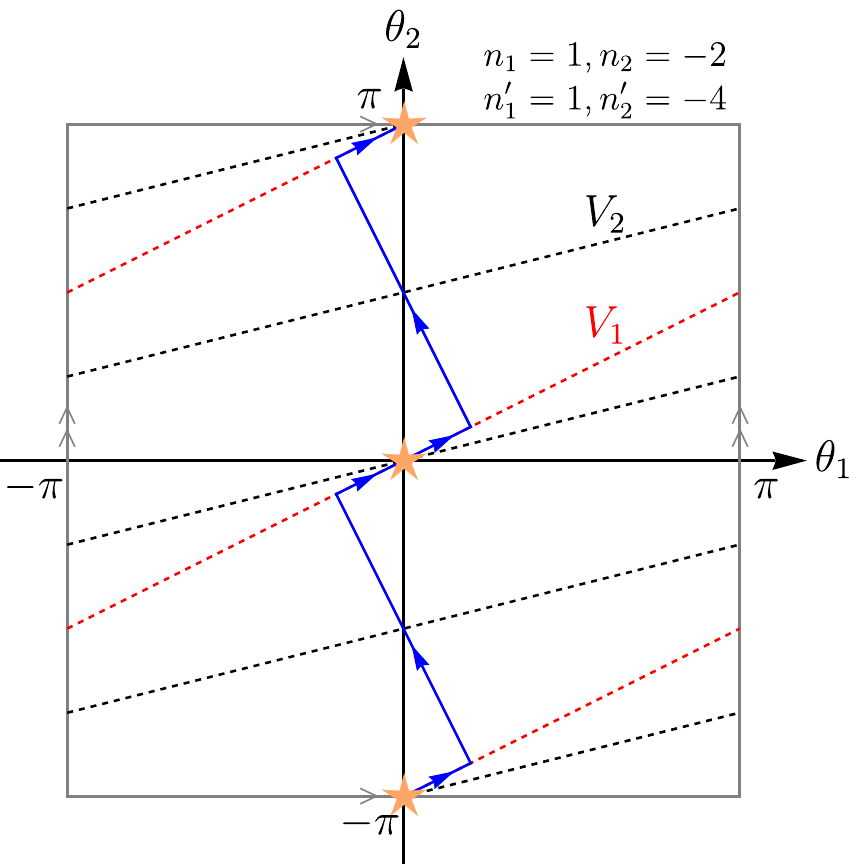}
    \end{center}
    \caption{%
        Same as the lower panel of Fig.~\ref{fig: pre-post} except for the choice of $(n_1', n_2') = (1, -4)$. 
        For illustrative purposes, we show $-\pi \leq \theta_1 \leq \pi$. The field trajectory (blue line) passes through two different minima (orange stars).
        }
    \label{fig: pre-post_D=2} 
\end{figure}

Let us first consider the case of $D\ne 0$. As an example case we adopt $(n_1', n_2') = (1, -3)$, which leads to $D = 1$.
In this case, $\phi_\mathrm{light}$ in each domain evolves so that $V_2$ takes its minimum value.
As a result, the DWs of $V_2$ are induced along those of $V_1$.
We show the field trajectory in this example case in the lower panel of Fig.~\ref{fig: pre-post}.
In each domain, the axions settle down to the potential minimum, $(\phi_1, \phi_2) = (0,0) \sim (2\pi f_1, 2\pi f_2)$. 
Between the domains, there is a DW of $V_1$, which is perpendicular to the red-dashed lines, and an induced DW of $V_2$, which is along the red-dashed line.
Consequently, the energy density of each domain becomes the same due to the induced DWs. Thus, the DW number of the $\phi_2$ string remains equal to $2$, and the string-wall network is stable.
For $D \geq 2$, DWs of $V_2$ are induced in a similar way.
In Fig.~\ref{fig: pre-post_D=2}, we show the field trajectory for $(n_1', n_2') = (1, -4)$, which leads to $D = 2$.
Thus, for $D \ne 0$, the DW number of $\phi_2$ string equals $|n_2| (\geq 2)$ even after $V_2$ becomes relevant, and the DW network remains stable.

\begin{figure}[!t]
    \begin{center}  
        \includegraphics[width=0.45\textwidth]{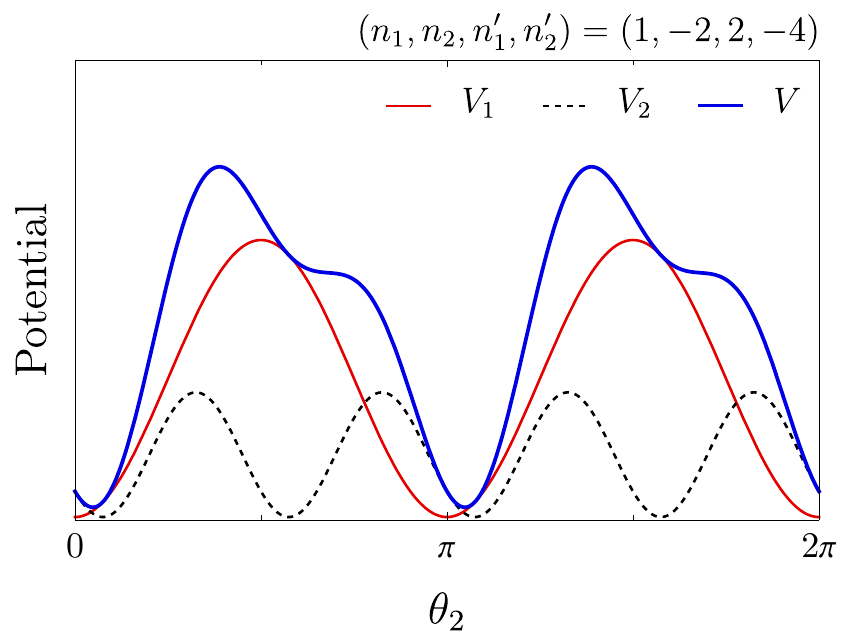}
    \end{center}
    \caption{%
        Dependence of the potentials on $\theta_2$ with fixed $\theta_1$ for $(n_1, n_2, n_1',n_2') = (1,-2,2,-4)$.
    }
    \label{fig: pre-post D=0 degenerate} 
\end{figure}

Next, we consider the case of $D = 0$. As an example we adopt $(n_1', n_2') = (2, -4)$, which leads to $D = 0$.
In this case, $\phi_\mathrm{light}$ remains massless even with $V_2$, since the potential depends on only $\phi_\mathrm{heavy}$.
We show the dependence of $V$ on $\theta_2$ with fixed $\theta_1$ in Fig.~\ref{fig: pre-post D=0 degenerate}.
Note that the relative phase $\alpha$ is generally nonzero in this case.
When we consider the potential around a string of $\phi_2$, i.e., $\theta_2 = 0$ to $2\pi$, there are two potential minima that are degenerate in energy, and thus the string-wall network is stable. In general, if $n_2'/n_2$ is an integer, there are $|n_2|$ degenerate minima around the $\phi_2$ string.

\begin{figure}[!t]
    \begin{center}  
        \includegraphics[width=0.45\textwidth]{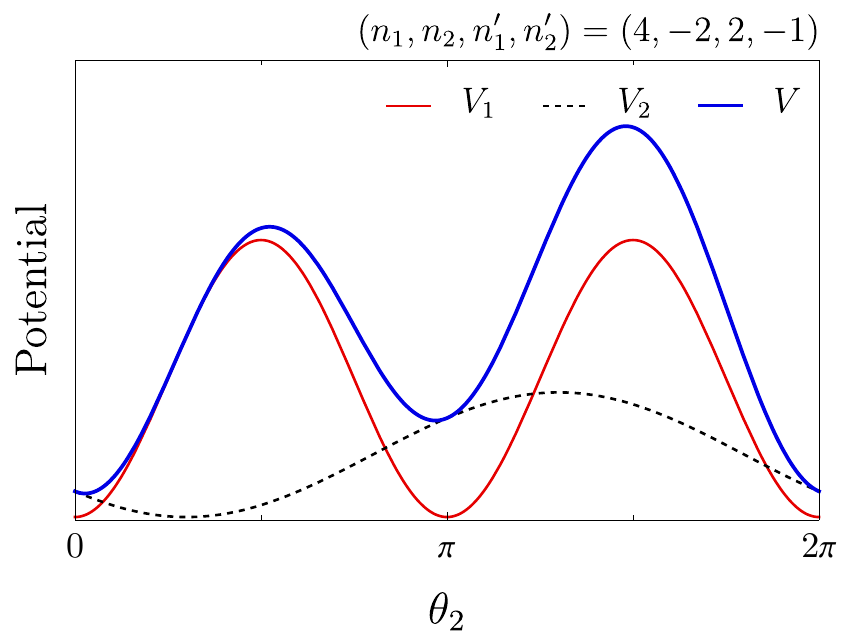}
    \end{center}
    \caption{%
        Dependence of the potentials on $\theta_2$ with fixed $\theta_1$ for $(n_1, n_2, n_1',n_2') = (4,-2,2,-1)$.
    }
    \label{fig: pre-post D=0 non-degenerate} 
\end{figure}

One can explicitly see that there is no such degeneracy if $n_2'/n_2$ is not an integer.
As an example, the case of $(n_1, n_2, n_1', n_2') = (4,-2,2,-1)$ is shown in Fig.~\ref{fig: pre-post D=0 non-degenerate}. In this case, the string-wall network collapses due to the potential bias between the domains induced by $V_2$.

So far, we have discussed the stability of the string-wall network assuming that both axions settle down to the potential minimum in each domain.
However, the axions do not start to roll down the potential until the Hubble parameter becomes comparable to the curvature of the potential.
In particular, after DWs are formed by $V_1$ but before $\phi_\mathrm{light}$ begins to move due to $V_2$, the difference in $V_2$ in each domain provides a potential bias.
The existence of the potential bias can already be seen in the upper panel of Fig.~\ref{fig: pre-post}.
When the DWs are formed by $V_1$, the configuration of the axion fields follows the blue line.
In particular, each domain corresponds to the blue line along with the minimum of $V_1$ shown by the red-dashed line.
The height of $V_2$ in each domain can be inferred from the distance of the blue line from the minimum of $V_2$ shown by the black-dashed lines.
Now, we see the height of $V_2$ in the two domains are different, which means the existence of the potential bias.
Since this bias disappears when $\phi_\mathrm{light}$ settles to the minimum of $V_2$ (see the lower panel of Fig.~\ref{fig: pre-post}), this potential bias is transient.
If the transient bias is sizable, the string-wall network collapses before $\phi_\mathrm{light}$ starts to oscillate or gains a population bias, which leads to the collapse of the string-wall network at later times.
Note that this transient bias is absent for the case of $D = 0$ because $V_2$ does not depend on $\phi_\mathrm{light}$ for $D = 0$.

\subsubsection{Post-post scenario}
\label{subsubsec: Second post-post}

Next, we consider the DW formation due to $V_2$ in the post-post scenario. As discussed in Sec.~\ref{subsubsec: first post-post}, when axions start to evolve due to $V_2$, there are two possibilities: either cosmic strings form string bundles in which DWs due to $V_1$ are confined, or the string-wall network extends throughout the entire space.

\begin{figure}[!t]
    \begin{center}  
        \includegraphics[width=0.45\textwidth]{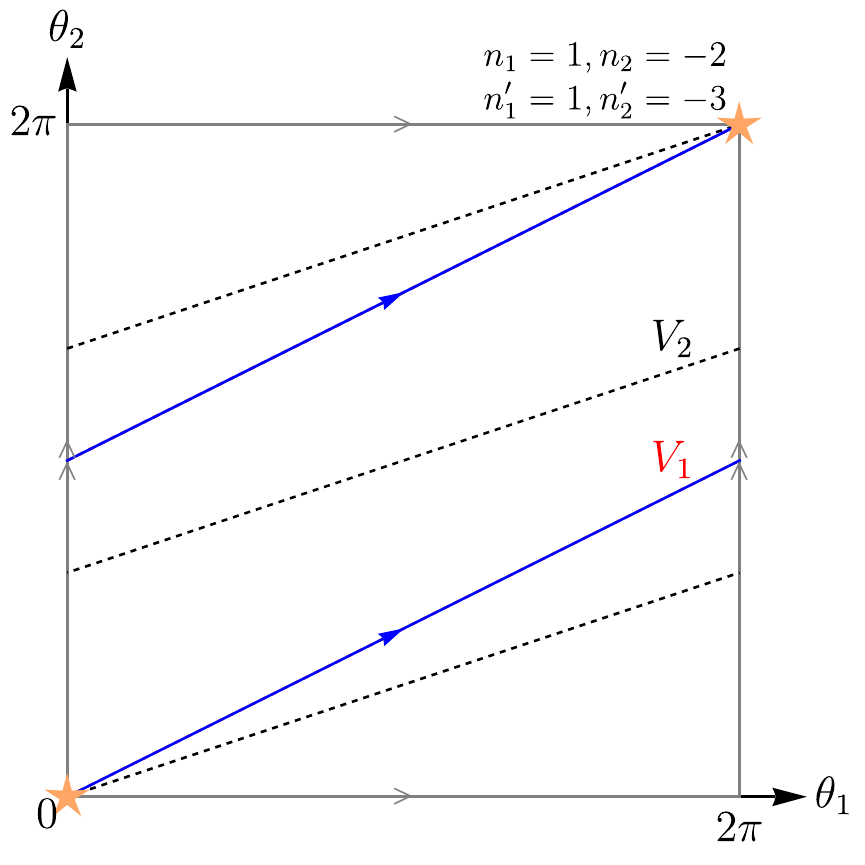}
    \end{center}
    \caption{%
        Field configuration of the axions around the string bundle in the post-post inflationary scenario with $N_\mathrm{DW} = 1$ is shown by the blue line.
        The winding numbers of $\phi_1$ and $\phi_2$ are $2$ and $1$, respectively.
        The trajectory is aligned with the minimum of $V_1$.
        The black-dashed line denotes the minimum of $V_2$, and the orange star is the unique potential minimum.
        Due to $V_2$, this string bundle has one DW.
    }
    \label{fig: post-post NDW=1} 
\end{figure}

Let us first consider the case when the string bundles are formed.
If $D \neq 0$, $V_2$ explicitly breaks the remaining $U(1)_L$ to $\mathbb{Z}_{N_{\rm DW}}$ with $N_\mathrm{DW} \equiv D/d$.
Then, $N_\mathrm{DW}$ DWs appear attached to each string bundle.
Consequently, the string-wall network collapses soon after $V_2$ becomes relevant for $N_\mathrm{DW} = 1$, while it remains stable for $N_\mathrm{DW} > 1$.
We show the trajectory of the axions around the string bundle for $N_\mathrm{DW} = 1$ in Fig.~\ref{fig: post-post NDW=1}. The DW due to $V_2$ is shown by the blue line, which overlaps with the minimum of $V_1$. The orange stars represent the identical potential minimum.

On the other hand, if $D = 0$, the $U(1)_L$ symmetry remains unbroken, and there is no DW attached to the string bundle.
Thus, the string bundles without DWs simply follow the scaling solution. Let us mention here a potential difference from the simple single-axion case.
Since the string bundles have an internal structure, it is nontrivial whether they evolve exactly in the same way as ordinary axion strings. 
For instance, we can consider a twisted loop formed by a string bundle.
If the twist of the loop is preserved, the contraction of the loop will stop at a certain radius.
Then, such twisted bundle loops are stable and may contribute to dark matter. 
However, we find that a string of one axion can easily pass through strings of the other axion in a lattice simulation.
As a result, the twisted loop is untwisted and decays similarly to a loop of a single string.
One way to prevent such untwisting is to introduce a repulsive interaction between the strings.
Further exploration of such twisted loops will be presented in a forthcoming paper.~\cite{Lee:preparation}

When the string-wall network extends throughout the entire space, the evolution of the network is very similar to the pre-post scenario.
If $D \neq 0$, $\phi_\mathrm{light}$ evolves due to $V_2$ in each domain, and induced DWs are formed as in the pre-post scenario.
There is then no potential bias between different domains, and the string-wall network remains stable, leading to the DW problem.
Note that a transient bias can arise before $\phi_\mathrm{light}$ evolves due to $V_2$, which may destabilize the DWs at later times.
On the other hand, if $D = 0$, the string-wall network is stable when $n_2'/n_2$ is an integer; otherwise, the network becomes unstable.

\subsection{Cases with one vanishing DW number}
\label{subsec: n1orn2=0}

Finally, we consider the case where one of the DW numbers vanishes. As mentioned in Sec.~\ref{sec: model}, we consider cases where either all  $n_1$, $n_2$, $n_1'$, and $n_2'$ are non-zero, or at most one of them is zero. Here we focus on the latter case.
For $n_1' = 0$ or $n_2' = 0$, the analysis presented above remains applicable.
We now turn to the cases where $n_1 = 0$ or $n_2 = 0$. 

\subsubsection{Post-post scenario with \texorpdfstring{$n_2 = 0$}{}}

In the post-post scenario, we consider $n_2 = 0$ without loss of generality. In this case we have $\phi_\mathrm{heavy} = \phi_1$ and $\phi_\mathrm{light} = \phi_2$.

For |$n_1| = 1$ and $n_2 = 0$, one DW of $V_1$ attaches to each cosmic string of $\phi_1$. Consequently, cosmic strings of $\phi_1$ rapidly decay due to the DW tension, while those of $\phi_2$ survive.
The subsequent dynamics then resembles that of a single axion model with a DW number of $n_2'$.

If $|n_1| \geq 2$ and $n_2 = 0$, a stable string-wall network forms due to $V_1$, where $|n_1|$ DWs are attached to each cosmic string of $\phi_1$.
This network remains stable at least until $V_2$ becomes relevant.
Given the hierarchy between $V_1$ and $V_2$, $\phi_1$ is fixed at the minimum of $V_1$ in each domain even when $V_2$ becomes relevant.

When $V_2$ becomes relevant, two phenomena occur. First, induced DWs of $V_2$ emerge along the existing DWs of $V_1$. Second, $|n_2'|$ DWs of $V_2$ form attached to each $\phi_2$ string.

The induced DWs of $V_2$ emerge because $\phi_1$ takes different values across the DWs of $V_1$. However, this is not the case when  $n_1'/n_1$ is an integer~\cite{Lee:2024xjb}.
This is because $V_2$ is minimized for the same value of $\phi_2$ between different domains separated by the DWs of $V_1$.

When $V_2$ becomes relevant, $\phi_2$ strings also develop $|n_2'|$ DWs of $V_2$.
Since the potential is minimized except for the interiors of strings and DWs due to the evolution of $\phi_2$, there is no potential bias between different domains separated by DWs of $V_2$. Thus, the string-wall network becomes stable for $|n_1| \geq 2$ and $n_2 = 0$. 

\subsubsection{Pre-post scenario with \texorpdfstring{$n_1 = 0$}{}}

In this case, the evolution of the string-wall network is almost the same as in the pre-post scenario discussed above.
The only difference is that the induced DWs are not formed if $n_2'/n_2$ is an integer.
See Ref.~\cite{Lee:2024xjb} for more detailed discussion.

\subsubsection{Pre-post scenario with \texorpdfstring{$n_2 = 0$}{}}

If $n_2 = 0$, $\phi_1$ coherently oscillates around the minimum of $V_1$.
Even after $V_2$ becomes relevant, $\phi_1$ is approximately fixed at the minimum of $V_1$.
Thus, the discussion of topological defects can be reduced to that of a single axion with the DW number of $|n_2'|$.

\subsection{Summary of classification}
\label{subsec: summary classification}

We summarize the classification of the fate of the string-wall network in the post-post and pre-post scenarios in Tables~\ref{tab: post-post} and \ref{tab: pre-post}, respectively. 
The columns labeled $V_1$ and $V_2$ in the upper row indicate what type of topological defect is formed or decays when each respective potential becomes effective.
Here ``$\phi_1$ network'' implies the string-wall network of $\phi_1$, and
``$\phi_2$ strings'' implies the $\phi_2$ string network.

\begin{table*}[htbp]
    \caption{Fate of the string-wall network in the post-post scenario. We consider cases where at most one of $n_1$, $n_2$, $n_1'$, $n_2'$ is zero. Here, we assume $|n_1| \geq |n_2|$ for convenience.
    For $|n_1| < |n_2|$, this table should read with $n_1 \leftrightarrow n_2$.
    $D \equiv |n_1 n_2' - n_2 n_1'|$. $d$ is the greatest common divisor of $(|n_1|,|n_2|)$, with $d = n_2$ for $n_1 = 0$ and $d = n_1$ for $n_2 = 0$. $N_\mathrm{DW} \equiv D/d$.
    Columns $V_1$ and $V_2$ show topological defects formed or decayed when each potential becomes effective.
    We assume that the effects of $V_1$ are more significant than those of $V_2$.}
    \label{tab: post-post}
    \begin{tabular}{ c | c | c || c | c | c | c  }
         $n_1$ & $n_2$ & $N_\mathrm{DW}$ & $V_1$ & $V_2$ & \multicolumn{2}{c}{figure}
        \\
        \hline \hline
        $1$ & $0$ & - & decay of $\phi_1$ strings & $\phi_2$ strings with $|n_2'|$ DWs & \multicolumn{2}{c}{-}
        \\
        \hline
        $\geq 2$ & $0$ & - & $\phi_1$ network \& $\phi_2$ strings & stable network & \multicolumn{2}{c}{-}
        \\
        \hline
        \multirow{3}{*}{$\geq 1$} & \multirow{3}{*}{$1$} & $0$ & \multirow{3}{*}{
             $\theta_L$ string bundles 
        } & $\theta_L$ string bundles 
            & \multirow{3}{*}{
            \begin{tabular}{c}
                 Figs.~\ref{fig: 2Dnumerical simulation}
                 and~\ref{fig: 3Dnumerical simulation}
            \end{tabular} 
            }
            & -
        \\
        \cline{3-3} \cline{5-5} \cline{7-7}
        & & $1$ & & decay of network & & Fig.~\ref{fig: post-post NDW=1}
        \\
        \cline{3-3} \cline{5-5} \cline{7-7}
        & & $\geq 2$ & & stable network & & -
        \\
        \hline
        $\geq 2$ & $\geq 2$ & - & stable network
         & 
         \begin{tabular}{c}
             stable network with induced DWs
             \vspace{-3pt}
             \\
             or collapse by transient bias 
        \end{tabular} 
        & Fig.~\ref{fig: 2Dnumerical simulation} & -
    \end{tabular}
\end{table*}

\begin{table*}[htbp]
    \caption{Fate of the string-wall network in the pre-post scenario. Other definitions and column descriptions are the same as in Table \ref{tab: post-post}.
    }
    \label{tab: pre-post}
    \begin{tabular}{ c | c | c || c | c | c }
        $|n_2|$ & $D$ & $n_2'/n_2$ & $V_1$ & $V_2$ & figure
        \\
        \hline \hline
        $0$ & - & - & $\phi_2$ strings without DWs & $\phi_2$ strings with $|n_2'|$ DWs & -
        \\
        \hline
        $1$ & - & - & decay of $\phi_2$ strings & no defects & -
        \\
        \hline
        \multirow{3}{*}{$\geq 2$} & \multirow{2}{*}{0} & integer & \multirow{3}{*}{\raisebox{-2\height}{$\phi_2$ strings with $|n_2|$ DWs}} & stable network & Fig.~\ref{fig: pre-post D=0 degenerate}
        \\
        \cline{3-3} \cline{5-6}
        & & non-integer 
        & & network collapse by bias & Fig.~\ref{fig: pre-post D=0 non-degenerate}
        \\
        \cline{2-3} \cline{5-6}
         & $\neq 0$ & - & & 
         \begin{tabular}{c}
             stable network with induced DWs\textsuperscript{*}
             \vspace{-3pt}
             \\
             or collapse by transient bias 
        \end{tabular}
         & Figs.~\ref{fig: pre-post} and \ref{fig: pre-post_D=2}
    \end{tabular}
    \footnotetext[1]{If $n_1 = 0$ and $n_2'/n_2$ is an integer, the stable network does not accompany induced DWs.}
\end{table*}

\section{Cosmological implications}
\label{sec: cosmology}

We have examined the evolution of the string-wall network by classifying the initial conditions and DW numbers to find that the network becomes stable in many cases.
In general, there is no reason for the minima of $V_1$ and $V_2$ to be aligned.
Thus, we expect $D \neq 0$ as a natural setup.
In other words, the string-wall network tends to be stable due to the formation of induced DWs, leading to the DW problem.

This tendency applies to models with $n$ axions and $n$ potential terms.
If the combinations of the axion fields in each potential are linearly independent, the axion fields settle to the global minimum of the total potential, and the string-wall network becomes stable.
To avoid the DW problem, we need to introduce an additional potential that works as a potential bias.
In this case, the string-wall network collapses when the pressure due to the potential bias overcomes the tension of the network.
Then, associated with the collapse of the string-wall network, axions and gravitational waves are emitted.
The spectrum of the emitted gravitational waves has a peak around the Hubble scale at the collapse, and its peak amplitude is estimated from the energy density of the string-wall network at that time~\cite{Hiramatsu:2013qaa}.
Since the network involves multiple species of strings and DWs, the typical frequency of the gravitational waves and the energy density of the string-wall network might be larger compared to the estimate based on simple $Z_2$ DW models.
Similarly, the string-wall network collapses when the energy density is larger than in the usual scenarios with a single axion. 
When the size of the false vacua, the string loops, or/and DWs
become comparable to the Schwarzschild radius, it could potentially collapse to form primordial black holes. Although the precise prediction of the abundance requires more detailed studies~\cite{Kitajima:2023cek} (see also Refs.~\cite{Vachaspati:2017hjw,Ferrer:2018uiu,Ge:2019ihf,Gelmini:2023ngs,Gouttenoire:2023ftk,Gouttenoire:2023gbn,Ferreira:2024eru,Dunsky:2024zdo}), 
our scenario can lead to an enhanced abundance of PBHs
due to the delayed collapse.

Alternatively, the string-wall network can collapse due to a transient bias that arises before the axions settle to the global minimum of the potential.
This transient bias may either directly collapse the network or indirectly destabilize the network by introducing a population bias.
As a result, the timescale of the string-wall network collapse can vary over a wide range.
Here, we briefly discuss the condition for the transient bias to be effective.
For simplicity, we assume that all the DW numbers are of $\mathcal{O}(1)$ and neglect them in the following.
We can estimate how the potential bias affects the DW evolution by comparing the energy of the DWs and potential bias in each Hubble volume.
The former is $\sim \sigma H^{-2}$ while the latter is $\sim \Lambda'^4 H^{-3}$.
Here, the tension of the DW of $V_1$, $\sigma$, is given by $\sim \Lambda^2 F$.
Thus, the transient bias is effective if it is present when $H \lesssim \Lambda'^4/(\Lambda^2 F)$.
Since $\phi_\mathrm{light}$ starts to oscillate when $H \simeq m_\mathrm{light} \sim \Lambda'^2/\sqrt{f_1^2+f_2^2}$, the transient bias is effective if $\Lambda'^2/\sqrt{f_1^2+f_2^2} \lesssim \Lambda'^4/(\Lambda^2 F)$, i.e., $f_1 f_2/(f_1^2+f_2^2) \lesssim \Lambda'^2/\Lambda^2$.
To realize such a situation, we need a hierarchy between $f_1$ and $f_2$ comparable to or more significant than that between $\Lambda^2$ and $\Lambda'^2$.

Lastly, we discuss a model with multiple axions where one combination behaves as the QCD axion. Up to this point, we have assumed that the axion potentials are time-independent. However, the potential of the QCD axion originates from non-perturbative QCD effects and thus has temperature dependence. Incorporating this temperature dependence into the analysis is straightforward.\footnote{The hierarchy of masses and tensions we have assumed so far may be altered due to the temperature dependence, potentially leading to interesting phenomena such as level crossing~\cite{Kitajima:2014xla,Daido:2015bva,Daido:2015cba,Ho:2018qur,Cyncynates:2023esj}.} Here, we focus on the impact of DWs of heavy axions, which is unique to multi-axion models.
For simplicity, let us consider a setup where the QCD axion corresponds to the lightest mass eigenstate. If string bundles are formed, the DWs of heavy axions are confined to the interior of the string bundles, and the system reduces to the usual single-axion model. However, more generally, for post-inflationary or mixed initial conditions, the effects of heavy axions cannot be integrated out, and DWs with high tension are typically formed. These DWs cause cosmological problems and must be collapsed.
One possibility, as mentioned earlier, is to introduce an additional potential to cause the collapse of the DW network. However, the introduction of such a bias term generally breaks the U(1) PQ symmetry explicitly beyond QCD, which spoils the high quality of U(1) PQ symmetry~\cite{Kamionkowski:1992mf, Holman:1992us, Barr:1992qq}. An alternative approach is to introduce an extra bias term in a direction orthogonal to the U(1) PQ symmetry. This approach allows for the collapse of the DW network without breaking the U(1) PQ symmetry. Therefore, to solve the strong CP problem with the QCD axion, we can either collapse the DWs without introducing additional bias terms, or carefully introduce a bias term orthogonal to the U(1) PQ symmetry.

One possibility along the former is a transient bias.  In this case, there is a possibility of collapsing the DWs without spoiling the high quality of the PQ symmetry. However, evaluating the lifetime of DWs given by such an induced population bias is non-trivial compared to the case of potential bias, requiring more detailed analysis.
On the other hand, in special cases where all potential terms have aligned minima, as in the $D = 0$ case discussed above, the string-wall network may collapse due to potential bias. However, in this case, the light axion decouples from QCD, preventing it from solving the strong CP problem.

Thus, the key features of the QCD axion model with multiple axions are the general existence of high-tension DW networks and the generation of gravitational waves accompanying their collapse. This is precisely what was pointed out in the context of the clockwork QCD axion in Ref.~\cite{Higaki:2016jjh}.\footnote{
Ref.~\cite{Higaki:2016jjh} considered post-inflationary conditions and stated that if $N_{\rm DW} = 1$, DWs would generally collapse if there were $N_{\rm axion}$ potential terms for $N_{\rm axion}$ axions. However, we have pointed out that this part is not correct, as the existence of induced walls was overlooked. } Here, we have seen that this holds true even in the case of multiple axions, particularly for two axions.

\section{Summary}
\label{sec: summary}

The QCD axion, a NG boson arising from the SSB of U(1) PQ symmetry, is currently a subject of intense research. While the structure of the PQ sector remains unknown, most studies on axion topological defects employ models with a single PQ scalar field. Our research investigates the structure and evolution of topological defects in scenarios with multiple PQ scalars, focusing particularly on the simplest case of two axions.

In single PQ scalar field models, initial conditions are limited to post-inflationary and pre-inflationary scenarios. However, the presence of multiple PQ scalar fields introduces the new possibility of mixed initial conditions combining pre- and post-inflationary conditions. We focused on two scenarios: the post-post scenario, where both axions have post-inflationary initial conditions, and the pre-post scenario, where one axion has a pre-inflationary initial condition and the other a post-inflationary one.

We introduced two hierarchical axion potential terms and examined the evolution of topological defects based on their DW numbers. In the post-post scenario, configurations of string bundles with heavy axions confined within them exist but form effectively only under specific initial conditions and DW numbers. More generally, we observe stable networks with high-tension DWs caused by heavy axions. This challenges the common assumption that heavy axions can be integrated out to simplify the system to a single-axion model, especially in the presence of cosmic strings. In the pre-post scenario, the formation of such string bundles is impossible due to one set of strings being diluted away by inflation, inevitably leading to DW networks. The ubiquity of these DW networks in multi-PQ scalar scenarios is a key argument of this paper, generalizing the observation made by \cite{Higaki:2016jjh} in the context of clockwork axions.

We summarized the classification of scenarios based on DW numbers in Tables \ref{tab: post-post} and \ref{tab: pre-post}. This classification shows a much greater complexity than single-axion models, including various phenomena like string bundles, induced DWs, and short-lived biases. Our findings suggest qualitative differences between multi- and single-axion models, challenging conventional perspectives on topological defects based on single-axion models.

We also discussed the cosmological implications of these scenarios, including the cosmological DW problem, gravitational wave generation, and applications to the QCD axion. For a viable cosmology, the string-wall network must collapse.
To this end, we may introduce an additional potential acting as a potential bias. However, when considering the QCD axion within a multi-axion system to solve the strong CP problem, introducing such additional potential terms compromises the quality of the QCD axion unless we carefully choose the direction of the bias term to be orthogonal to the U(1) PQ symmetry.
Therefore, in the case of the QCD axion, DW collapse through transient bias is favorable. 
In any case, the collapse of high-tension string-wall networks results in copious emission of axions and gravitational waves. While the spectra can be roughly estimated based on the collapse timing, differences from single-axion models require detailed numerical simulations.

The key message of this paper is that multiple PQ scalars significantly alter the evolution of topological defects, with important implications for axion and gravitational wave production. Indeed, in the domain of axion physics, more is different.

\section*{Acknowledgments}
This work is supported by JSPS Core-to-Core Program (grant number: JPJSCCA20200002) (F.T.), JSPS KAKENHI Grant Numbers 20H01894 (F.T.), 20H05851 (F.T. and W.Y.), 21K20364 (W.Y.), 22H01215 (W.Y.), 22K14029 (W.Y.), 23KJ0088 (K.M.), and 24K17039, (K.M.), Graduate Program on Physics for the Universe (J.L.) and JST SPRING Grant Number JPMJPS2114 (J.L.),  Incentive Research Fund for Young Researchers from Tokyo Metropolitan University (W.Y.). 
This article is based upon work from COST Action COSMIC WISPers CA21106, supported by COST (European Cooperation in Science and Technology).

\appendix

\section{String bundle formation}

\begin{figure}[htbp]
    \begin{center}  
        \includegraphics[width=0.45\textwidth]{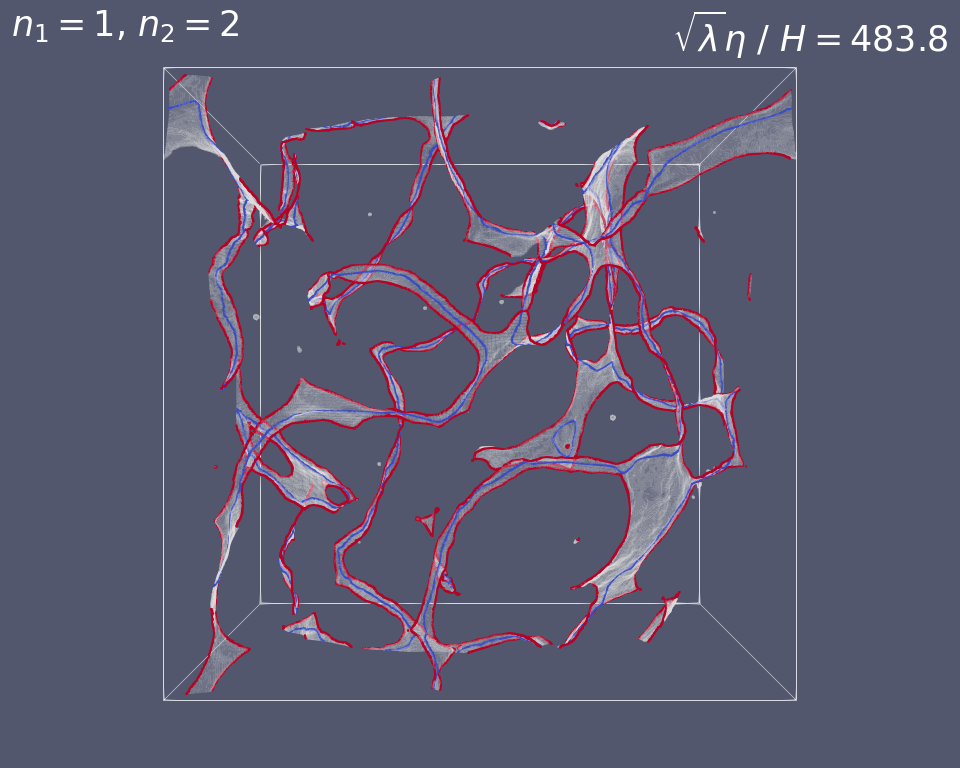}
    \end{center}
    \caption{
    The snapshots of strings and DWs at the time when $H^{-1} = 483.8/\sqrt{\lambda}\eta$ for $(n_1,n_2) = (1,2)$ are shown. 
    The blue (red) lines correspond to 1-strings (2-strings) and the semitransparent white surfaces correspond to heavy axion DWs.
    One side of the box is the size of two Hubble horizons.
    }
    \label{fig: 3Dnumerical simulation} 
\end{figure}

We have discussed the heavy axion DW formation and the subsequent string bundle formation when $|n_1|=1$ and $|n_2|={\cal O}(1)$ (or vice versa) in the post-post scenario in Sec.~\ref{subsubsec: first post-post}.
Once the DWs are formed, they start to dominate the dynamics of strings in their scaling regime, and they will contract within ${\cal O}(1)$ Hubble time producing string bundles.  This has been confirmed with 2D simulations (see Fig.~\ref{fig: 2Dnumerical simulation}).

Here, we show the numerical results from 3D lattice simulations to confirm that our expectations for the string bundle formation hold when three-dimensional structures are included, and thus the late-time dynamics of defects converges to that of a single-axion scenario.

\begin{figure}[ht]
    \begin{center}  
        \includegraphics[width=0.45\textwidth]{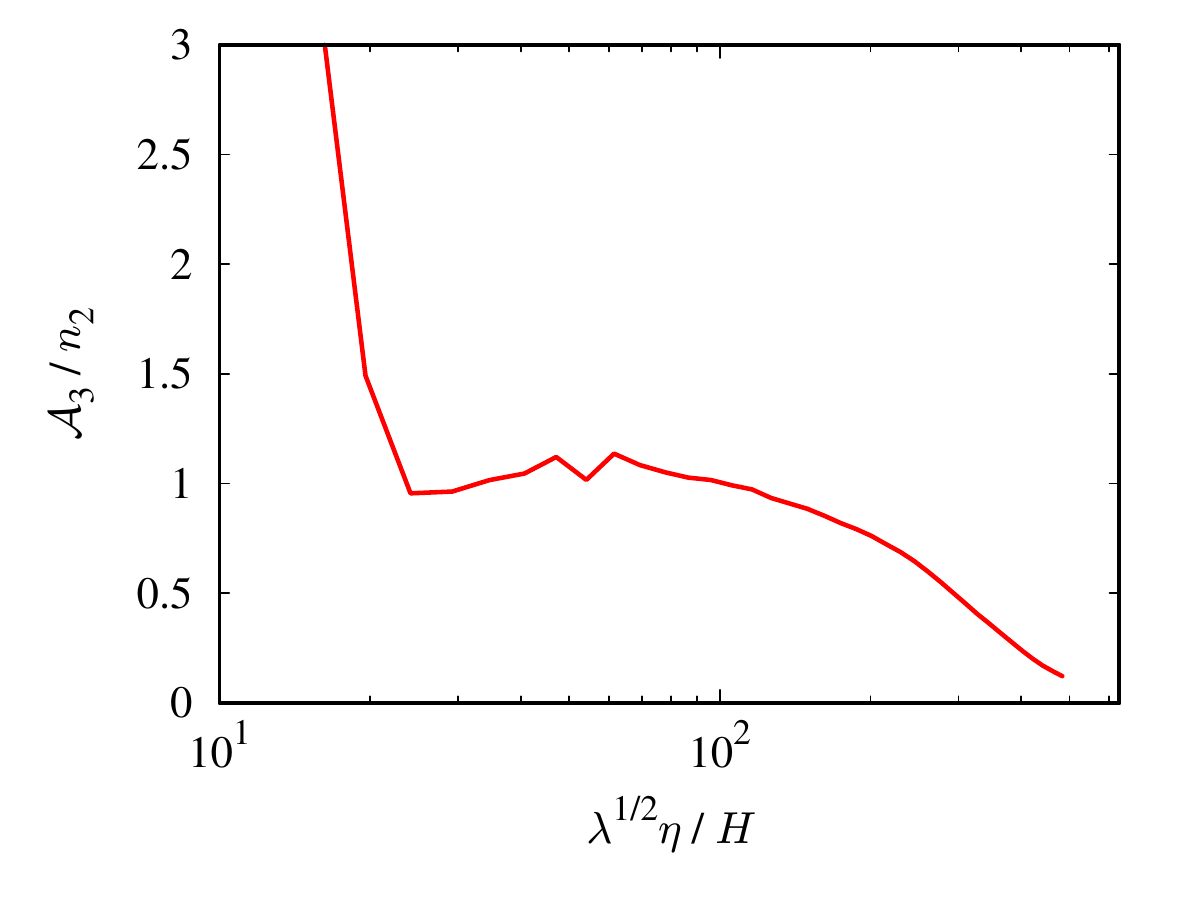}
    \end{center}
    \caption{
    The time evolution of DW area parameter ${\cal A}_3$ normalized by $n_2$.
    Note that the strings are formed and start to follow their scaling solutions at $2t=H^{-1} \simeq 60 / \sqrt{\lambda}\eta$, and two scaling solutions of the strings become different each other due to the confinement of the DWs after $H^{-1} \simeq 90 / \sqrt{\lambda}\eta$.
    The area parameter decreases and asymptotes to zero as the string bundles are being formed.
    }
    \label{fig: evolve_dw_area_n1=1_n2=2} 
\end{figure}

We set $n_1=1$, $n_2=2$, $f_1 = f_2 \equiv \sqrt{2}\eta$, and $\lambda_1 = \lambda_2 \equiv \lambda$.
The mass of $\phi_{\rm heavy}$ is set to be $m_{\rm heavy}^2 = 0.2 \lambda \eta^2$.
We start the simulation from the time when $H^{-1} = 1/\sqrt{\lambda}\eta$ with random field values for each site.
Fig.~\ref{fig: 3Dnumerical simulation} shows a snapshot of the spatial distribution of strings and DWs in the 3D lattice.
Note that the snapshot contains $2^3$ Hubble horizons.
Most neighboring strings are aligned, and all the visibly remaining DWs have sizes smaller than the Hubble horizon, so we expect the string bundles to form soon.
The shrinking of the DWs can be seen in Fig.~\ref{fig: evolve_dw_area_n1=1_n2=2}, which plots the time evolution of the area parameter of the DWs, ${\cal A}_3$, normalized by $n_2$.
The area parameter indicates the number of DWs per one Hubble horizon patch, and the detail of the definition is the same as in Ref.~\cite{Kitajima:2023kzu}; we count the lattice sites where the field value $\phi_{\rm heavy}$ crosses the potential hilltop.

Here, we comment on the junction structures of $\phi_{\rm light}$ string networks, which can be seen in the snapshot. The junctions would be a possible deviation from the dynamics of the single-axion string, if they continued to exist.
However, these junctions are considered to be unstable, because they disappear once pair annihilations or interconnections of strings occur.
Even in the presence of interactions between the strings, they easily pass through each other due to the motion induced by the string tension. Thus, the junctions are only transient objects, and do not affect the overall time evolution of the string network.

\bibliographystyle{JHEP}
\bibliography{ref}

\end{document}